# Comparison Research of Millimeter-Wave/Infrared Co-aperture Reflector Antenna Systems Based on a Specialized Film

1st Zongze Li
*Beijing Key Laboratory of Millimeter wave and Terahertz Techniques*
*Beijing Institute of Technology*
Beijing, China
3220231850@bit.edu.cn

2nd Xinlong Yang
*Beijing Key Laboratory of Millimeter wave and Terahertz Techniques*
*Beijing Institute of Technology*
Beijing, China
3220242124@bit.edu.cn

3rd Yiming Zhao
*Advanced Research Institute of Multidisiplinary Sciences*
*Beijing Institute of Technology*
Beijing, China
zhaoyiming@bit.edu.cn

4th Chen Yao
*School of Information and Electronics*
*Beijing Institute of Technology*
Beijing, China
chenyao@bit.edu.cn

5th Hongda Lu
*Beijing Key Laboratory of Millimeter wave and Terahertz Techniques*
*Beijing Institute of Technology*
Beijing, China
luhongda@bit.edu.cn

6th Yong Liu
*Beijing Key Laboratory of Millimeter wave and Terahertz Techniques*
*Beijing Institute of Technology*
Beijing, China
fatufo@bit.edu.cn

*Abstract*—This paper presents a novel co-aperture reflector antenna operating in millimeter-wave (MMW) and infrared (IR) for cloud detection radar. The proposed design combines a back-fed dual-reflector antenna, an IR optical reflection system, and a specialize thin film with IR-reflective/MMW-transmissive properties. Simulations demonstrate a gain exceeding 50 dBi within 94 GHz ± 500 MHz bandwidth, with 0.46° beamwidth in both azimuth (E-plane) and elevation (H-plane) and sidelobe levels below -25 dB. This co-aperture architecture addresses the limitations of standalone MMW and IR radars, enabling high-resolution cloud microphysical parameter retrieval while minimizing system footprint. The design lays a foundation for airborne/spaceborne multi-mode detection systems.

*Keywords—common-aperture, millimeter-wave, infrared, W-band, cloud detection*

## I. Introduction

Millimeter-wave cloud radars [1] offer strong capability for detecting cloud and fog micro-particles, high spatial resolution with narrow beams even under compact aperture sizes, and the ability to continuously retrieve cloud microphysical parameters. However, these radars suffer significant attenuation during precipitation and limited horizontal coverage. In contrast, infrared laser cloud radars [2] excel in identifying cloud base height and phase, though their effectiveness is limited in studying cloud microphysics.

A combination of lidar and millimeter-wave radar can leverage the advantages of both, enabling microphysical retrievals and continuous detection even under rainy conditions. In most existing dual-mode systems, millimeter-wave and infrared signals are processed separately, resulting in bulky and costly systems. A co-aperture solution is needed to resolve these challenges.

Recent works have explored co-aperture antenna designs. Studies [3] introduced central apertures in flat arrays to accommodate infrared/laser sensors behind the array, achieving co-aperture integration. However, such central cutouts can degrade the millimeter-wave antenna performance—larger apertures worsen side-lobes, while smaller ones compromise optical performance. In [4], a single-reflector structure is used for millimeter-waves and a Cassegrain configuration for infrared, with a shared main reflector to achieve co-aperture. Yet, the secondary reflector in the optical path obstructs the millimeter-wave feed, degrading the side-lobe performance. These designs generally exhibit side-lobe levels around -15 dB and limited gain due to aperture size constraints, failing to meet cloud radar requirements for high gain and low side-lobe performance. Thus, a new composite antenna architecture is needed. This paper presents a novel millimeter-wave/infrared co-aperture composite antenna tailored for cloud-profiling radars. The system incorporates a thin film that reflects infrared and transmits millimeter waves, placed above the secondary reflector, avoiding the need for a perforated main reflector and performance trade-offs between the two systems.

## II. Antenna Design

### A. Scheme Selection

In millimeter-wave/infrared co-aperture integration schemes, the feed illumination configuration can be typically categorized as back-fed or offset-fed.

The back-fed configuration offers a lower profile, with coaxial reflectors enabling easier co-aperture integration of infrared lasers. Two integration approaches are proposed (Fig. 1):

(a) A 45°-tilted infrared mirror, coated for microwave transmission and infrared reflection, is mounted between the feed and sub-reflector. The mirror is secured in a glass plate attached to the feed via a flange. A side-mounted laser reflects off the mirror toward the sub-reflector, then to the main reflector, sharing the millimeter-wave path.(b) A Cassegrain optical design places the mirror above the sub-reflector, utilizing it as the primary optical surface.

This work was supported in part by the National Natural Science Foundation of China under Grant 62271047, and Grant 12173006.

The offset-fed design eliminates sub-reflector blockage, improving millimeter-wave performance (Fig. 2). Here, a horizontal mirror directs the laser (mounted above) through the sub-reflector and main reflector for co-aperture operation.

Trade-offs exist: rear-fed suffers from sub-reflector blockage (worse sidelobes), while offset-fed has larger lateral dimensions (increased weight). The optimal choice depends on simulation results below.

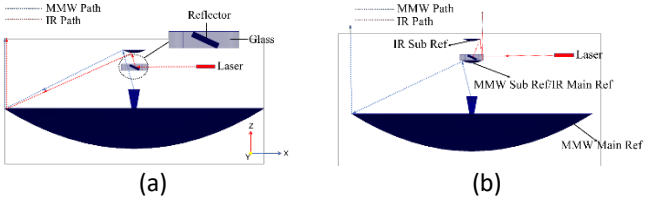

Fig. 1. (a) Back-fed configuration 1,(b) Back-fed configuration 2.

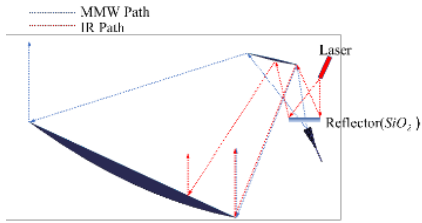

Fig. 2. Offset-fed configuration.

### B. Dual Polarized Horn Feed

For reflector antennas, under the condition of fixed reflector geometry and excluding special cases such as feed defocusing, the overall antenna performance is primarily determined by the feed system. Consequently, feed design becomes one of the critical aspects.

Based on the previously determined reflector configuration and radar performance requirements, the feed must meet the following specifications: excellent beam rotational symmetry, dual-polarization capability with high port isolation. To satisfy these design criteria, a corrugated conical horn antenna (as illustrated) was developed, comprising three main sections: the input waveguide section, mode transition section, and output section.

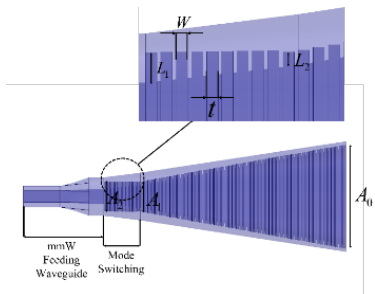

Fig. 3. Horn features.

The simulated results of VSWR and radiation patterns are presented below.

As illustrated in Fig. 5, the feed exhibits a voltage standing wave ratio (VSWR) below 1.35 across the entire 75-110 GHz frequency band. Within the 93.0-95.5 GHz sub-band, the E-plane and H-plane main beams demonstrate nearly identical patterns within the ±15° angular region.

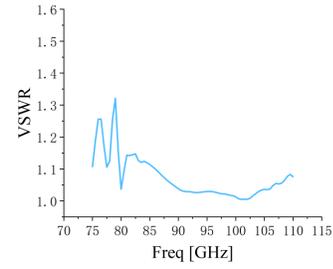

Fig. 4. Simulated VSWR.

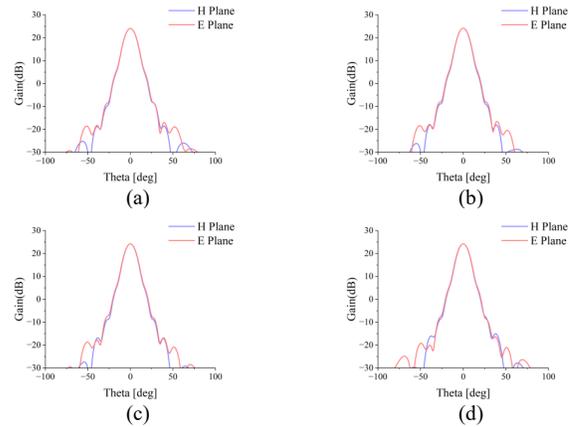

Fig. 5. Simulated radiation pattern. (a) 93 GHz,(b) 94 GHz,(c) 95 GHz, (d) 96 GHz.

To enable dual-polarization in the single-input feed, an ortho-mode transducer (OMT) must connect to the square waveguide, requiring high isolation, strong cross-polarization suppression, and low loss. $Port_{in1}$/$Port_{in2}$ handle horizontal/vertical polarizations respectively, with Portout linked to the waveguide.

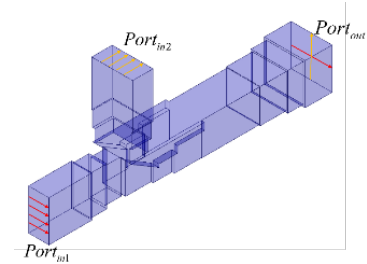

Fig. 6. OMT features.

The performance characteristics of the ortho-mode transducer (OMT) are summarized as follows.

Input return loss is above 16dB,inter-port isolation is below -48dB,insertion loss is below 0.1dB and cross polarization is below -52dB.

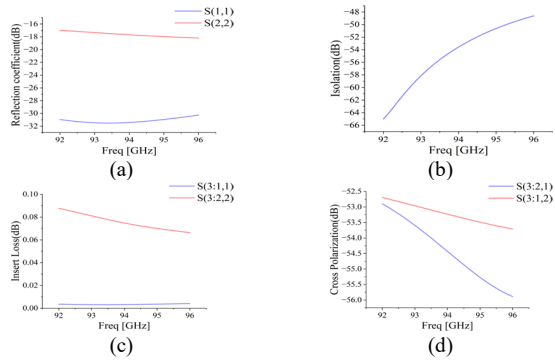

Fig. 7. S parameters.(a) Reflection coefficient,(b) Isolation,(c) Insertion Loss, (d) Cross Polarization.

## C. Simulation Results of Reflector Antenna System

As evidenced by the preceding analysis, structural considerations alone prove insufficient for determining the optimal solution, necessitating a comprehensive evaluation of each configuration's electromagnetic performance.

The simulation results for the offset-fed configuration are presented as follows.

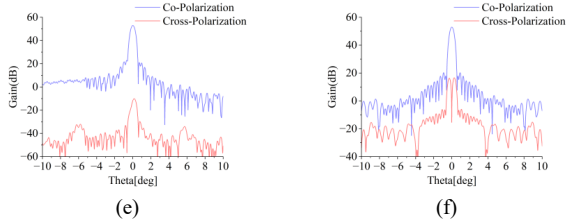

Fig. 8. Simulated radiation pattern.(a) Phi = 0°,(b) Phi = 90°.

As shown in Fig. 8, both E-/H-planes achieve gain=52.97 dB with sidelobes=-30.08/-32.71 dB, beamwidths=0.43°/0.41°, and cross-polarization=-73.08/-32.7 dB, meeting design requirements. However, the scheme has engineering drawbacks: large lateral size, high weight (350 kg), and transport/testing challenges due to marble-based optical stability demands.

The simulation results of the back-fed scheme (a) are as follows.

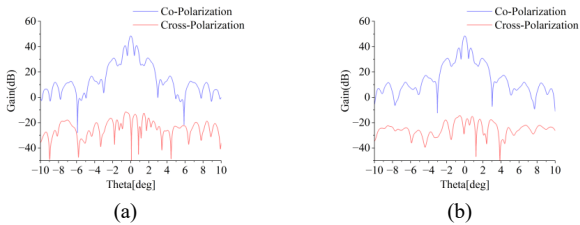

Fig. 9. Simulated radiation pattern.(a) Phi = 0°,(b) Phi = 90°.

As shown in the figure, both the E-plane and H-plane achieve a gain of 48.42 dB, with sidelobe levels of -7.61 dB and -9.44 dB, beamwidths of 0.36° and 0.38°, and cross-polarization levels of -59.97 dB and -62.87 dB, respectively. Compared to the offset-fed scheme, this design exhibits a noticeable degradation in performance, particularly in terms of gain and sidelobe levels.

The simulation results of the back-fed scheme (b) are as follows.

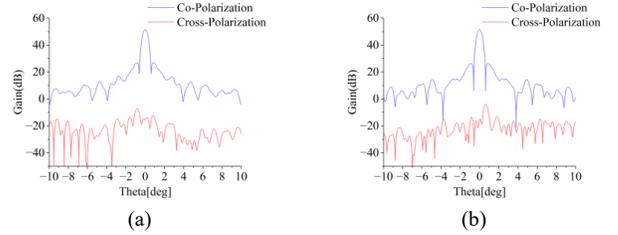

Fig. 10. Simulated radiation pattern.(a) Phi = 0°,(b) Phi = 90°.

As shown in the figure, both the E-plane and H-plane achieve a gain of 51.46 dB, with sidelobe levels of -25.07 dB and -25.18 dB, identical beamwidths of 0.47°, and cross-polarization levels of -58.46 dB and -55.46 dB, respectively. These results indicate that this design represents the optimal solution. The final schematic is shown in Fig 11.

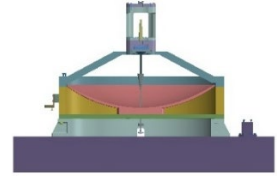

Fig. 11. Section View.

## III. CONCLUSION

This study proposes a millimeter-wave/infrared co-aperture antenna for cloud-probing radar, integrating back-fed reflectors and infrared optics to achieve dual-modal sensing. The design delivers >50 dBi gain, 0.46° beamwidth, and <-25 dB sidelobes at 94 GHz ± 500 MHz, enhancing resolution and penetration. The optimized back-fed configuration achieves 51.46 dB gain with symmetric beams and compact structure, suitable for airborne/spaceborne systems.